\documentclass[emulateapj,apj,dvipdfmx]{emulateapj}
\newcommand{\ts}{t_{\rm stop}}
\newcommand{\cs}{c_{\rm s}}
\newcommand{\cd}{c_{\rm d}}
\newcommand{\sd}{\Sigma_{\rm d}}

\newcommand{\Hd}{H_{\rm d}}

\shorttitle{An Origin of Multiple Ring Structure in HL Tau}
\shortauthors{Takahashi \& Inutsuka}

\usepackage{natbib, aas_macros}
\begin{document}
\title{
An Origin of Multiple Ring Structure and Hidden Planets  in HL Tau: A Unified Picture by Secular Gravitational Instability}

\author{Sanemichi Z. Takahashi\altaffilmark{1}, Shu-ichiro Inutsuka\altaffilmark{2}}
\altaffiltext{1}{Astronomical Institute, Tohoku University,6-3 Aoba, 
Aramaki-aza, Aoba-ku, Sendai, Miyagi,Japan;
sanemichi@astr.tohoku.ac.jp}
\altaffiltext{2}{Department of Physics, Nagoya University, Furo-cho, 
Chikusa-ku, Nagoya, Aichi, 464-8602, Japan;
inutsuka@nagoya-u.jp
}

\begin{abstract}
Recent ALMA observation has revealed multiple ring structures formed in a protoplanetary disk around HL Tau.
Prior to the ALMA observation of HL Tau, theoretical analysis of secular gravitational instability (GI) described a possible formation of multiple ring structures with separations of 13 AU around a radius of 100 AU in protoplanetary disks under certain conditions.  
In this article, we reanalyze the viability of secular GI by adopting the physical values inferred from the observations. 
We derive the radial distributions of the most unstable wavelength and the growth timescale of secular GI and verify that secular GI can form the ring structures observed in HL Tau.
When a turbulent viscosity coefficient $\alpha$ remains small in inner region of the disk, secular GI grows in the whole disk.
Thus, the formation of planetary mass objects should occur first in the inner region as a result of gravitational fragmentation after the non-linear growth of secular GI.
In this case, resulting objects are expected to create the gaps at $r\sim 10$ AU and $\sim 30$ AU.
As a result, all ring structures in HL Tau can be created by secular GI.
If this scenario is realized in HL Tau, the outer region corresponds to the earlier growth phase of the most unstable mode of secular GI, and  the inner region corresponds to the outcome of the non-linear growth of secular GI.  Therefore, this interpretation suggests that we are possibly witnessing both the beginning and end of planet formation in HL Tau.
\end{abstract}


\def\bm#1{\mbox{\boldmath $#1$}}

\section{INTRODUCTION}
Planets are expected to form in protoplanetary disks.
Since the disks affect the planet formation processes, it is important to understand the formation and evolution of protoplanetary disks.
Although there are many numerical simulations of the  formation of protoplanetary disks (e.g. \cite{1998ApJ...508L..95B,2007ApJ...670.1198M,2010ApJ...714L..58T,2010ApJ...718L..58I,2011MNRAS.416..591T};  see also reviews by \cite{2012PTEP.2012aA307I,2016PASA...33...10T}), long term evolution of the disks is still unclear.
On the other hand, high-angular-resolution direct imaging of protoplanetary disks recently became available.
The observations have revealed that non-axisymmetric structure \cite[e.g.][]{2013Sci...340.1199V,2013Natur.493..191C,2013PASJ...65L..14F}, and ring like structures \cite[e.g.][]{2007A&A...469L..35G,2010ApJ...725.1735I,2012ApJ...747..136I,2013ApJ...775...30I,2011ApJ...732...42A,2011ApJ...729L..17H,2012ApJ...758L..19H,2012ApJ...753...59M,2012ApJ...760L..26M} are formed in protoplanetary disks.
These structures may provide clues to understand an evolution of protoplanetary disks and formation processes of planets.

Recently, the result from the high-resolution observation of the HL Tau region with ALMA was reported by \cite{2015ApJ...808L...3A}.
HL Tau is a well-known T Tauri star and has been extensively observed.
It is deeply embedded in circumstellar gas and observed as a reflection nebula at optical wavelength \cite[]{1995ApJ...449..888S}.
The disk-like infalling envelope whose radius is about 1400 AU and the accretion disk whose radius is about 150 AU are observed around HL Tau in $^{13}$CO and infrared emission \cite[]{1993ApJ...418L..71H,1997ApJ...478..766C}, and the outflow is also observed as the optical image \cite[]{1988ApJ...333L..69M}.
The ALMA observation of HL Tau has revealed that multiple ring structures have been formed in the disk around HL Tau \cite[]{2015ApJ...808L...3A}.
The observed ring structure of the brightness temperature and surface density structures estimated from the observation is shown in Figure 3 in \cite{2015ApJ...808L...3A} and Figure 3 in \cite{2016ApJ...816...25P}, respectively.
This stimulated extensive studies to find physical mechanisms to form the multiple ring structure of HL Tau; gap opening due to planets \cite[]{2015ApJ...806L..15K,2015MNRAS.453L..73D,2016ApJ...818..158A}, dust growth \cite[]{2015IAUGA..2256118Z}, baroclinic instability \cite[]{2015MNRAS.453L..78L}, and the growth and radial drift of dust grains with the effect of sintering \cite[]{2016ApJ...821...82O}, but formation mechanisms of such structures remain unclear. 
Prior to the ALMA observation of HL Tau, \cite{2014ApJ...794...55T} (hereafter TI14) proposed a possible formation of multiple ring structures by secular gravitational instability \cite[secular GI, cf.][]{2000orem.book...75W, 2011ApJ...731...99Y, 2011ApJ...738...73S, 2010ApJ...719.1021M,2016ApJ...817..140S}.
Secular GI forms ring structures whose width is about 10AU in timescale $10^{4 \-- 5}$ yr at about 100AU.
Weak turbulence and enhanced dust-to-gas mass ratio are required for secular GI.
TI14 indicates that multiple ring structures may be an indicator of the dust concentration and weak turbulence in the disk.
Detailed analysis of the result of HL Tau observation reveals that weak turbulence (a turbulent viscosity coefficient $\alpha \sim$ a few $10^{-4}$) is required to reproduce the ALMA observation \cite[]{2016ApJ...816...25P}. 
\cite{2016ApJ...816...25P} have estimated the parameter $\alpha$ from the dust scaleheight.
The dust scaleheight is given by the balance between the dust settling and diffusion by turbulence in the vertical direction.
Since the disk of HL Tau is inclined, the contrast of the dust continuum emission decreases on the minor axis when the dust scaleheight is large.
Thus, \cite{2016ApJ...816...25P} proposed that 
small dust scaleheight that corresponds to $\alpha \approx 3\times 10^{-4}$ is required to reproduce the observed contrast on the minor axis of HL Tau disk\footnote{
\cite{2016ApJ...816...25P} has pointed out that $\alpha$ obtained from the dust scale height is different from that estimated from the mass accretion rate ($\alpha \sim 10^{-2}$).
This discrepancy may suggest that the disk of HL Tau is not steady, or the angular momentum of the disk is transferred by the mechanisms other than the turbulence, for example, magnetic braking \cite[]{2013ApJ...769...76B,2013ApJ...772...96B,2014ApJ...784..121S}.
When the disk is not steady, the accretion rate onto the central star can be different from the accretion rate at $r=100$~AU,
and when the angular momentum is transferred by the mechanisms other than the turbulence, the accretion rate does not reflect the strength of the turbulence.
Therefore, the parameter $\alpha$ estimated from the gas accretion rate is not necessarily related to the turbulence at $r=100$~AU.
On the other hand, $\alpha$ estimated from the dust scaleheight is directly related to the turbulence in the disk.
Since the turbulence that causes the dust diffusion stabilize the secular GI, we use $\alpha$ obtained from the dust scaleheight in this work.}.
Moreover, the disk of HL Tau is gravitationally unstable if we assume typical dust-to-gas mass ratio $\epsilon =0.01$ \cite[]{2011ApJ...741....3K,2015ApJ...808..102K}. When disks become gravitationally unstable, spiral arms are formed in the disk.
However, spiral structures are not observed in HL Tau.
Since observation only gives the dust surface density of the disk, gas surface density decreases as adopted dust-to-gas mass ratio increases.
Therefore, the observation suggests that the disk is gravitationally stable but dust-to-gas mass ratio is larger than typical value $\epsilon =0.01$.
These two features are consistent with what TI14 have anticipated.
Therefore, secular GI is a promising mechanism of ring structure formation in HL Tau.
In this work, we perform a linear stability analysis of secular GI adopting the physical values obtained from observation as the backgrounds state.
We obtain the most unstable wavelength and growth timescale of secular GI and discuss that secular GI can form the ring structures observed in HL Tau.

This paper is organized as follows. Basic equations for secular GI are given in Section \ref{basic_equations}. 
In Section \ref{comparison}, we adopt the observational results of HL Tau as background values and derive the dispersion relation and the radial distribution of the maximum wavelength and the growth timescale of secular GI. We compare these results with the observation of ring structures of HL Tau.
We discuss 
a possible unified scenario for the formation of both rings and planets
in HL Tau caused by secular GI in Section \ref{discussion}. 
A summary is given in Section \ref{conclusion}.

\section{BASIC EQUATIONS}
\label{basic_equations}
In this section, we show basic equations for a linear stability analysis of secular GI.
We use two fluid equations for both gas and dust.
We focus on purely horizontal motions of gas and dust, and use vertically integrated equations of continuity and motion for both gas and dust, and the Poisson equation.
We take into account turbulent viscosity of gas, diffusion of dust caused by turbulence of gas, and velocity dispersion of dust.
\begin{equation}
 \frac{\partial \Sigma}{\partial t} + \bm{\nabla}\cdot(\Sigma
  \bm{u})=0,
\end{equation}
\begin{eqnarray}
\lefteqn{\Sigma \left(\frac{\partial u_i}{\partial t}+u_k \frac{\partial u_i}{\partial x_k} \right) =}
\nonumber \\ 
&-&c_s^2\frac{\partial \Sigma}{\partial x_i}-\Sigma\frac{\partial}{\partial x_i}\left(\Phi - \frac{GM_*}{r}\right) + \frac{\Sigma_d(v_i-u_i)}{t_{\rm stop}} \nonumber \\ 
&+&\frac{\partial}{\partial x_k} \left[\Sigma \nu \left(\frac{\partial u_i}{\partial x_k} +\frac{\partial u_k}{\partial x_i}
-\frac{2}{3}\delta_{ik}\frac{\partial u_l}{\partial x_l}\right)\right].
\end{eqnarray}
\begin{equation}
  \frac{\partial \Sigma_d}{\partial t} + \bm{\nabla}\cdot(\Sigma_d
   \bm{v})=D\nabla ^2\Sigma,
\end{equation}
\begin{eqnarray}
\lefteqn{\sd\left(\frac{\partial \bm{v}}{\partial
t}+(\bm{v}\cdot\bm{\nabla})\bm{v}\right)} \nonumber\\
&=&-\cd^2\nabla\sd
-\Sigma_d\bm{\nabla}\left(\Phi
- \frac{GM_*}{r}\right)
+\frac{\Sigma_d(\bm{u}-\bm{v})}{t_{\rm stop}}.
\end{eqnarray}
\begin{equation}
 \nabla^2\Phi=4\pi G(\Sigma+\Sigma_d)\delta(z),
\end{equation}
where $\Sigma$ and $ {\bm u}$ are surface density and velocity of gas, 
$\sd$ and ${\bm v}$ are surface density and velocity of dust,
$\cs$ is the sound speed of gas, 
$\nu$ is the coefficient of kinematic viscosity caused by turbulence, 
$D$ is the diffusivity of the dust
due to the gas turbulence,
$\cd$ is the velocity dispersion of the dust,
$M_*$ is the central star mass, 
$\Phi$ is the gravitational potential of the gas and dust, 
and 
$\ts$ is the stopping time of a dust particle.

We adopt a local shearing box model
and the local radial and azimuthal coordinates are
$(x,y)$, which rotate with 
the Keplerian frequency $\Omega$
\cite[e.g.][]{1965MNRAS.130..125G,1987MNRAS.228....1N}.
Since we investigate the ring structure formation, we assume axisymmetry for simplicity.
We assume a steady state background  with uniform surface density;
$\Sigma_0 , \Sigma_{{\rm d}0} = {\rm const}$;
dust-to-gas mass ratio $\epsilon=\Sigma_{{\rm d}0}/\Sigma_0$;
and Keplerian rotation
$u_{x0}=v_{x0}=0,u_{y0}=v_{y0}=(-3/2)\Omega x$.
We decompose the physical quantities into background values and small
perturbations proportional to $\exp[ikx-i\omega t]$.
The linearized equations are given as follows:
\begin{equation}
 -i\omega \delta \Sigma + ik \Sigma_0 \delta u_x=0,
\label{eq:eoc_gas_l}
\end{equation}
\begin{eqnarray}
-i\omega  \delta u_x -2\Omega \delta u_y &=&
 -\cs^2\frac{ik\delta\Sigma}{\Sigma_0 } - ik \delta\Phi + \frac{\epsilon
 (\delta v_x- \delta u_x)}{t_{\rm stop}} \nonumber\\
&&-\frac{4}{3}\nu k^2\delta u_x,
\label{eq:eom_gas_x_l_vis}
\end{eqnarray}
\begin{equation}
-i\omega \delta u_y+\frac{\Omega}{2} \delta u_x=
\frac{\epsilon (\delta v_y-\delta u_y)}{t_{\rm stop}}
-\nu k^2 \delta u_y -ik\frac{3\nu\Omega}{2\Sigma_0}\delta \Sigma.
\label{eq:eom_gas_y_l_vis}
\end{equation}
\begin{equation}
  -i\omega \delta \sd + ik \epsilon \Sigma_0  \delta v_x=-Dk^2 \delta \sd,
\label{eq:eoc_dust_l}
\end{equation}

\begin{equation}
-i\omega \delta v_x -2\Omega \delta v_y =
 -\cd^2 \frac{ik \delta \sd}{\Sigma_0} -ik\delta \Phi
+ \frac{\delta u_x-\delta v_x}
{t_{\rm stop}}.
\label{eq:eom_dust_x_l_p}
\end{equation}
\begin{equation}
-i\omega \delta v_y +\frac{\Omega}{2}  \delta v_x =
\frac{\delta u_y-\delta v_y}{t_{\rm stop}},
\label{eq:eom_dust_y_l}
\end{equation}
\begin{equation}
 \delta \Phi=- \frac{2\pi G (\delta \Sigma+\delta \sd )}{|k|}.
\label{eq:poisson_l}
\end{equation}
We take into account the effect of the thickness of the gas and dust on
the gravitational potential and rewrite Equation (\ref{eq:poisson_l})
\cite[]{1970ApJ...161...87V, 1984prin.conf..513S}:
\begin{equation}
\delta \Phi = - 2\pi G\left(\frac{\delta \Sigma}{1+kH}+
\frac{\delta \sd}{1+ k \Hd}\right),
\label{eq:poisson_l_disk_height}
\end{equation}
where $H =\cs/\Omega$ and $\Hd$ are the scale height of the gas and dust.
The relation between $\Hd$ and $H$ is given as follows
\cite{2007Icar..192..588Y}:
\begin{equation}
 \Hd =
  H\left(1+\frac{\ts\Omega}{\alpha}\frac{1+2\ts\Omega}{1+\ts\Omega}\right)
^{-1/2}.
\end{equation}
When $\alpha \ll \ts\Omega \ll 1$,
\begin{equation}
  \Hd \simeq
  \left(\frac{\ts\Omega}{\alpha}\right)
^{-1/2}H.
\end{equation}
The relation between the turbulent viscosity and dust diffusivity is given
by \cite{2007Icar..192..588Y} \cite[see also][]{2010ApJ...719.1021M}:
\begin{equation}
 {\bar D} = \frac{1+\ts\Omega
  +4(\ts\Omega)^2}{[1+(\ts\Omega)^2]^2}\alpha,
\end{equation}
where ${\bar D}\equiv D \Omega/\cs^2$ is the normalized dust diffusivity  
and $\alpha \equiv \nu \Omega/\cs^2$ is the dimensionless measure of
turbulent intensity \cite[]{1973A&A....24..337S}.
In a turbulent disk, the velocity dispersion of dust is estimated by 
\cite{2007Icar..192..588Y}:
\begin{equation}
 \cd^2 =  \frac{1+2\ts\Omega +(5/4)(\ts\Omega)^2}{[1+(\ts\Omega)^2]^2}          \alpha\cs^2. 
  \label{eq:cd}
\end{equation}
Hereafter we use the growth rate of the instability $n\equiv -i\omega$ instead of the frequency $\omega$.
For given values of $\epsilon$, $\alpha$, $\ts\Omega$, and Toomre parameter $Q\equiv \cs\Omega/(\pi G \Sigma_0)$ \cite[]{1964ApJ...139.1217T}, we obtain normalized dispersion relations ($n/\Omega$ as a function of $kH$) from Equations (\ref{eq:eoc_gas_l}) to  (\ref{eq:cd}).

\section{COMPARISON WITH THE OBSERVATION OF HL TAU}
\label{comparison}
In this section, we calculate the most unstable wavelength and the growth timescale of secular GI.
We adopt physical values obtained from the observation of HL Tau as the background values of linear stability analyses.
We assume that turbulence in HL Tau is not strong and adopt $\alpha = 3\times 10^{-4}$ as observationally suggested by \cite{2016ApJ...816...25P},
This small value of turbulence strength is also suggested by recent theoretical modeling of magnetorotational instability \cite[]{2016ApJ...817...52M}. We discuss this aspect in Section \ref{alpha_distribution}.

We use two models of dust surface density and temperature distributions proposed by previous work.
The gas surface density is not obtained from the observation.
However, if we assume dust-to-gas mass ratio $\epsilon=0.01$, the disk becomes gravitationally unstable since the dust surface density is expected to be very large to reproduce the observed values.
If the disk is gravitationally unstable and $Q\lesssim 2$ is satisfied, the spiral arms should be formed in the disk.
However, the observation of HL Tau shows that no spiral arm is formed in the disk.
In this work, we adopt constant $\epsilon$ for each models so that $Q > 2$ is satisfied in the disk.
The dust radius is also an important parameter since stopping time $\ts$  given by Epstein drag law is proportional to the dust radius: $\ts = \rho_{\rm int} a /(\rho_{\rm g}\cs)$, where $\rho_{\rm int}$ is the internal density of dust, $a$ is the dust radius, and $\rho_{\rm g}$ is gas density.
In \cite{2016ApJ...816...25P} the maximum and minimum dust sizes are 3~mm and 0.03~${\rm \mu m}$, and the power of the dust size distribution is -3.5 (integrated over the whole disk).
Since the dust mass of the disk is dominated by the largest dust, we simply adopt $a=3$~mm in this work\footnote{
\cite{2016ApJ...816...25P} have mentioned that the power of the dust size distribution in outer $\sim 80$ AU of the disk is about -4.5 and the mm size dust is depleted there.
Although we assume $a=3$~mm for simplicity in this work, further analysis of the observational data and investigation of secular GI including the effect of dust distribution are required \cite[see, e.g.][]{2016ApJ...817..140S}.
}.
The central star mass is obtained by \cite{2015ApJ...808L...3A}: $M_{\rm *} = 0.7 M_{\rm \odot}$.

\subsection{Exponentially Cutoff Disk Model}
\cite{2015ApJ...808..102K} obtain the dust surface density distribution by fitting CARMA data as follows:
\begin{equation}
 \sd(r)=0.51 \left(\frac{r}{R_{\rm c}}\right)^{-\gamma}
  \exp\left[-\left(\frac{r}{R_{\rm c}}\right)^{2-\gamma}\right]
  \ [\rm g \ cm^{-2}],
  \label{eq:dust_surface_exp_cutoff}
\end{equation}
where $\gamma = -0.2$, $R_{\rm c} = 80.2$ AU.
\cite{2016ApJ...816...25P} shows that this dust surface density distribution approximately reproduces the ALMA observation data.
The temperature distribution at the midplane obtained in \cite{2015ApJ...808..102K} is about half of the brightness temperature distribution given by \cite{2015ApJ...808L...3A}.
Thus, we adopt the following temperature distribution:
\begin{equation}
 T(r) = 30 \left(\frac{r}{20\ [\rm AU]}\right)^{-0.65}\ [\rm K].
\label{eq:Tr_ALMA}
\end{equation}
These dust surface density and temperature distribution give $\sd\approx 0.11 \ {\rm g\ cm^{-2}}$, and $T\approx 11$ K at 100 AU. We assume $\epsilon = 0.02$ for this model.
Hereafter, we call this model ``exponentially cutoff disk model''.
In this case, the disk is gravitationally stable ($Q\approx 2.9$ at 100 AU).
We obtain $\ts\Omega \approx 0.42$ by using dust radius $a=3$ mm, $\rho_{\rm int}=3\,{\rm g\, cm^{-3}}$ and $\rho_{\rm gas} = \Sigma/(\sqrt{2\pi}H)$, where $\Sigma = \Sigma_{\rm d}/\epsilon $ is gas surface density.

\begin{figure}[t]
 \epsscale{1}
 \includegraphics[width=8cm]{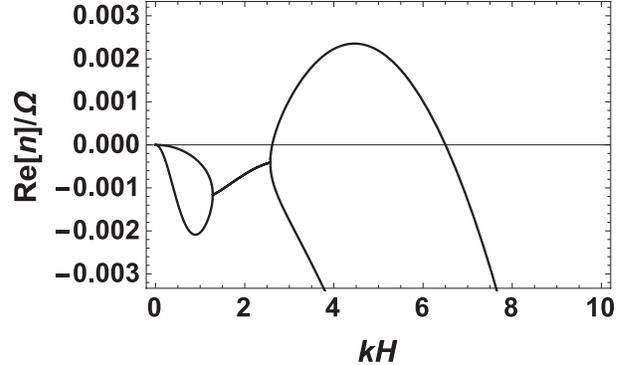}
\caption{
Dispersion relation of secular GI at 100 AU for exponentially cutoff disk model.
The horizontal axis is the normalized wavenumber, $kH$, and the vertical axis is the normalized growth rate of the instability, ${\rm Re}[n]/\Omega$.}
\label{fig:disp_exp_cut}
\end{figure}

Fig. \ref{fig:disp_exp_cut} shows the dispersion relation of secular GI at 100 AU. 
The most unstable wavelength is about 11 AU. It agrees with the observation of HL Tau.
The growth timescale of secular GI, 
$t_{\rm grow} = ({\rm Re}[n])^{-1}$, is about $9\times 10^4$ yr. 
Since HL Tau is thought to be young, $t_{\rm grow} < 10^6$yr is required, but long growth timescale is required to observe the growing unstable mode. 
In this sense, the result $t_{\rm grow} \sim 10^5 $ yr is consistent with the scenario that observed  ring structures are formed by secular GI.

\begin{figure}[t]
\includegraphics[width=8cm]{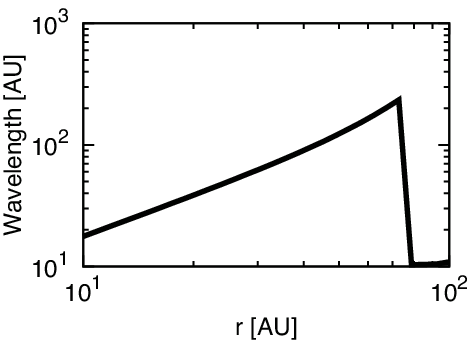}
\includegraphics[width=8cm]{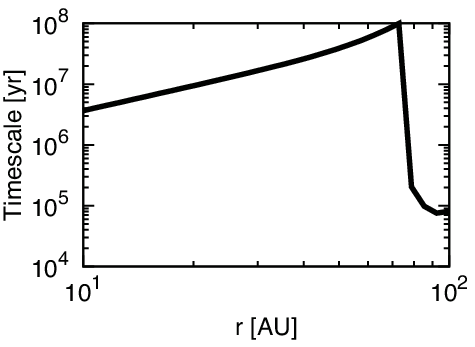}
 \caption{The distribution of the most unstable wavelength (upper panel) and the growth timescale (lower panel) for exponentially cutoff disk model.
The horizontal axis is the radial distance from the central star.
The vertical axis of the upper panel is the most unstable wavelength and the lower panel is the growth timescale.
Secular GI grows in the region $r\gtrsim 80$ AU.
In the region $r\lesssim 80$ AU, secular GI does not grow but viscous overstability grows.
}
\label{fig:lambda_tgrow_exp_cutoff}
\end{figure}

Fig. \ref{fig:lambda_tgrow_exp_cutoff} shows radial distributions of the most unstable wavelength and the growth timescale for exponentially cutoff disk model.
Secular GI grows in the region where the radius $r$ is larger than about 80 AU. The most unstable wavelength is about 10 AU and the growth time scale is about $10^5$ yr in this region. 
Therefore, a few rings observed in the region $r \gtrsim 80$ AU can be created by secular GI.
In the region $r\lesssim 80$ AU, secular GI is stable but viscous overstability grows \cite[cf.][]{1995Icar..115..304S}. 
However, since the growth timescale is larger than the typical disk lifetime and the most unstable wavelength is longer than the radius, viscous overstability does not grow in the disk. In this work, we neglect viscous overstability and focus on secular GI.

\begin{figure}[t]
\includegraphics[width=8cm]{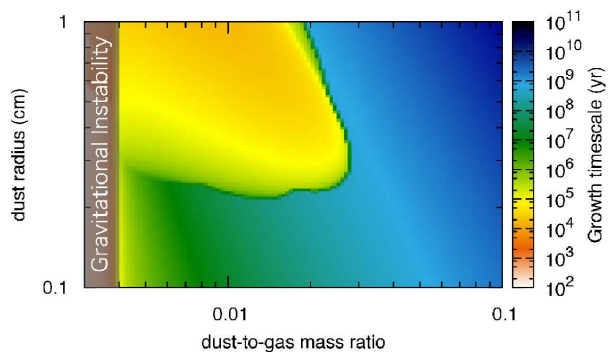}
 \caption{
Growth timescale of the instability for various $\epsilon$ and $a$.
The horizontal axis is the dust-to-gas mass ratio $\epsilon$ and the vertical axis is dust radius $a$.
Secular GI grows in the region shown by yellow and red.
For $\epsilon \lesssim 4\times 10^{-3}$, gravitational instability of the gas disk grows.
}
\label{fig:e-a-tgrow}
\end{figure}

It is difficult to obtain the dust radius and the surface density of gas from the observation. 
Thus, there is large uncertainty in the model parameters $a$ and $\epsilon$.
Fig. \ref{fig:e-a-tgrow} shows the growth timescale of the instability for various $\epsilon$ and $a$.
Secular GI grows in the region shown by yellow and red.
This figure shows that $a\gtrsim 2$ mm and $\epsilon \lesssim 0.03$ are required for secular GI in this model.
For $\epsilon \lesssim 4\times 10^{-3}$, gravitational instability of the gas disk grows. The criterion $\epsilon = 4\times 10^{-3}$ corresponds to $Q\approx 0.6$, which is smaller than the critical $Q$ value for the razor-thin disk because of the disk thickness. 

In Section \ref{discussion}, we discuss a scenario in which all the rings in the disk can be created by secular GI

\subsection{Power-law Disk Model}
\begin{figure}[t]
\includegraphics[width=8cm]{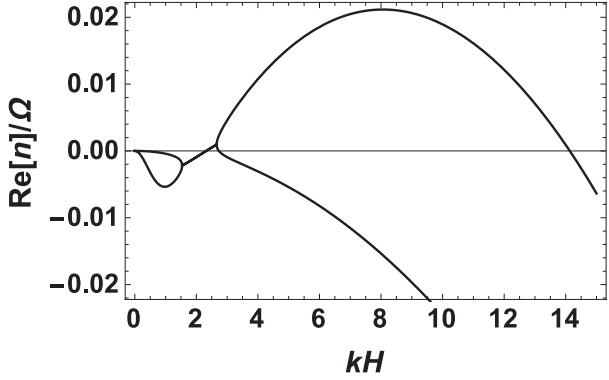}
\caption{
Dispersion relation of secular GI at 100 AU for power-law disk model.
The horizontal axis is the normalized wavenumber, $kH$ and the vertical axis is the normalized growth rate of the instability, ${\rm Re}[n]/\Omega$.
}
\label{fig:disp}
\end{figure}

Since the dust opacity depends on the details of the dust grain property, for example the shape of dust, the physical values obtained from the observation should have some uncertainties.
Therefore, it is worth investigating the instability with various disk models.
For this purpose, we used the surface density and temperature distributions given by \cite{2016ApJ...820...54K} as an unperturbed state in this section.
In \cite{2016ApJ...820...54K}, ALMA Band 6 continuum along the major axis is fitted by a power-law temperature and dust surface density profile with depression (gaps) at the gap radius.
This model is obtained from only the observed intensity of Band 6 of ALMA and does not fit Band 3 and 7 data. 
Although this model may not be adequate for the surface density and the temperature profile of HL Tau, 
we use this model without depression as background state to investigate whether secular GI can grow in another background state.
The dust surface density without depression is given by Equation (4) in \cite{2016ApJ...820...54K}: $\Sigma_d = 8.3 (r/1[\rm AU])^{-0.3} {\rm \ g\ cm^{-2}}$. 
This surface density represents the observed intensity at the bright rings.
In this work, however, we have to use the surface density before the ring formation as a background state of secular GI. 
Thus, we overestimate the dust surface density when we adopt $\Sigma_d = 8.3 (r/1[\rm AU])^{-0.3} {\rm \ g\ cm^{-2}}$ as a background state, because this dust surface density should correspond to the maximum dust surface density after secular GI grows and concentrates the dust.
To avoid overestimating the dust surface density, we halve the dust surface density:
\begin{equation}
 \sd = 4.15 \left(\frac{r}{1\ [\rm AU]}\right)^{-0.3} [\rm g\ cm^{-2}]\label{eq:sigma_d_prof}
\end{equation}
\begin{equation}
 T = 280 \left(\frac{r}{1\ [\rm AU]}\right)^{-0.3} [\rm K]\label{eq:T_prof}
\end{equation}
We assume $\epsilon = 0.1$ for this model. 
Hereafter, we call this model ``power-law disk model''.
In this case, other parameters are given by $Q\approx 3.8$ and $\ts\Omega \approx 0.22$ at 100 AU.
Fig. \ref{fig:disp} shows dispersion relation of Secular GI at 100AU.
The most unstable wavelength is about 16 AU and growth timescale are about $9 \times 10^3$ yr.
Thus, ring structures formed by secular GI also agree with observation when we adopt the power-law disk model.

\begin{figure}[t]
\includegraphics[width=8cm]{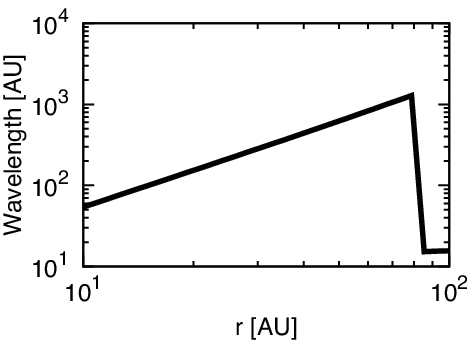}
\includegraphics[width=8cm]{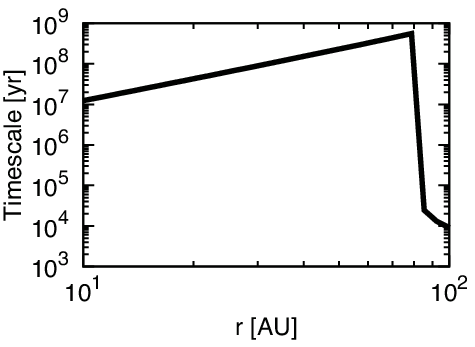}
 \caption{The distribution of the most unstable wavelength (upper panel) and the growth timescale (lower panel) for power-law disk model.
Secular GI grows in the region $r\gtrsim 80$ AU.
}
\label{fig:r-lambda}
\end{figure}

Fig. \ref{fig:r-lambda} shows the radial distribution of the most unstable wavelength and growth timescale for power-law disk model.
Secular GI grows in the region $r \gtrsim 80$ AU but cannot grow in $r \lesssim 80$ AU.
Therefore, either in the exponentially cutoff or power-law disk model, rings in the region $r\lesssim 80$ AU are not supposed to be created by in-situ secular GI.
We need other disk models to reproduce all ring structures in the disk by  secular GI, which are discussed in a following section.

\section{DISCUSSION}
\label{discussion}
\subsection{Strength of Turbulence in the Disk}
\label{alpha_distribution}
One of the most interesting features of HL Tau is that turbulence seems to be weaker than that usually expected.
The ALMA observation of HL Tau indicates that the coefficient of turbulent viscosity $\alpha$ is of order $10^{-4}$ in the disk.
However,
$\alpha$ due to MRI turbulence is usually expected to be large ($\sim 10^{-2}$) \cite[e.g.,][]{1995ApJ...440..742H,2004ApJ...605..321S}.
One candidate of a physical mechanism to suppress the MRI turbulence is electron heating introduced by \cite{2015ApJ...800...47O}. \cite{2016ApJ...817...52M} suggests that MRI is suppressed by electron heating and $\alpha \lesssim 10^{-3}$ can be realized in protoplanetary disks.
When electron heating suppresses the MRI turbulence, $\alpha$ is expected to be small in inner region of the disk.
Figure 9 in \cite{2016ApJ...817...52M} suggests $d\log\alpha/d\log r \gtrsim 2$ at the midplane.
If $\alpha$ is smaller than $3\times 10^{-4}$ in the region $r\lesssim 80$ AU, the region where secular GI grows is larger than that obtained in Section \ref{comparison}.

In this section, we discuss the secular GI in the disk in which $\alpha$ increases with radius as expected from electron heating in contrast to the previous section in which we used the constant $\alpha$ disk model following \cite{2016ApJ...816...25P}.
We try to model small $\alpha$ in inner region by the following ad hoc formula;
\begin{equation}
 \alpha_1 = 3\times 10^{-4}\left(\frac{r}{100~{\rm AU}}\right)^2,
\end{equation}
\begin{equation}
 \alpha_2= 3\times10^{-4}\left(\frac{r}{100~{\rm AU}}\right)^3.
\end{equation}

\begin{figure}[t]
\includegraphics[width=8cm]{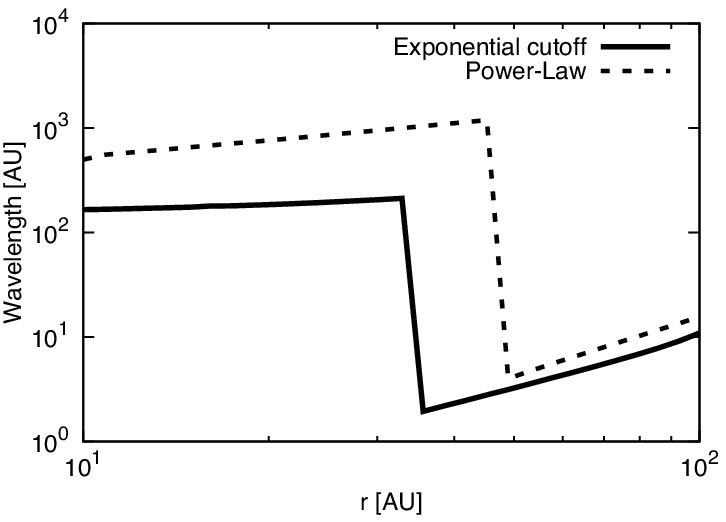}
\includegraphics[width=8cm]{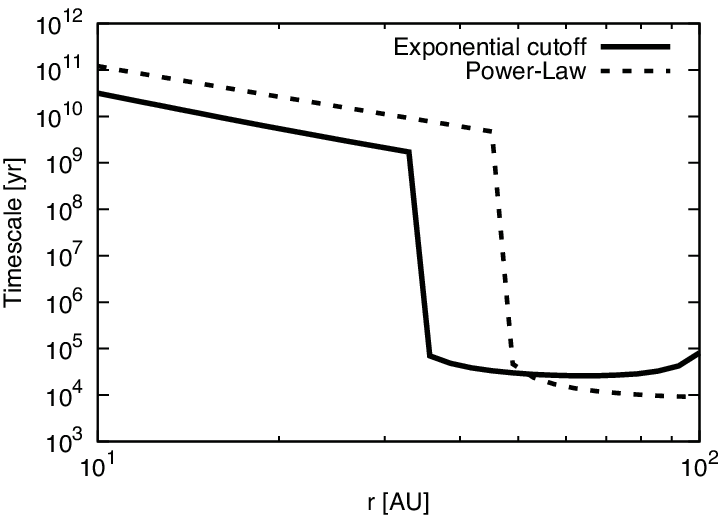}
 \caption{The distribution of the most unstable wavelength (upper panel) and the growth timescale (lower panel) for exponentially cutoff disk model (solid line) and power-law disk model (dotted line) with $\alpha = 3\times 10^{-4} (r/100\,[\rm AU])^2$.
Secular GI grows in the region $r\gtrsim 40$ AU for exponentially cutoff disk model and $r\gtrsim 50$ AU for power-law disk model.
}
\label{fig:lambda_tgrow_alpha2}
\end{figure}

\begin{figure}[t]
\includegraphics[width=8cm]{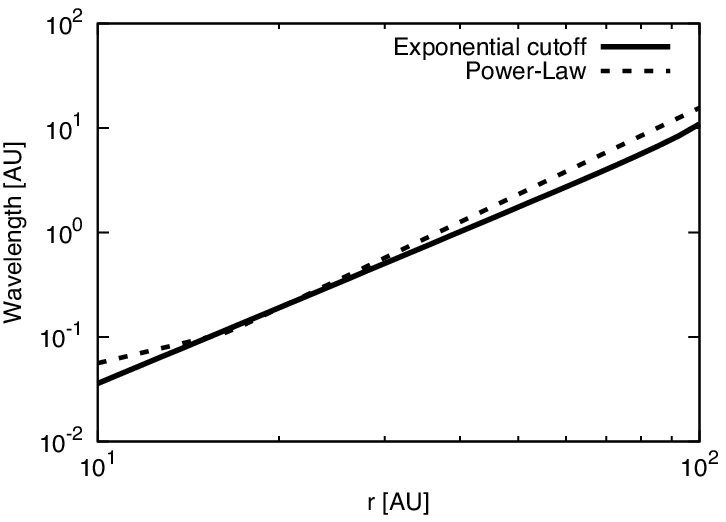}
\includegraphics[width=8cm]{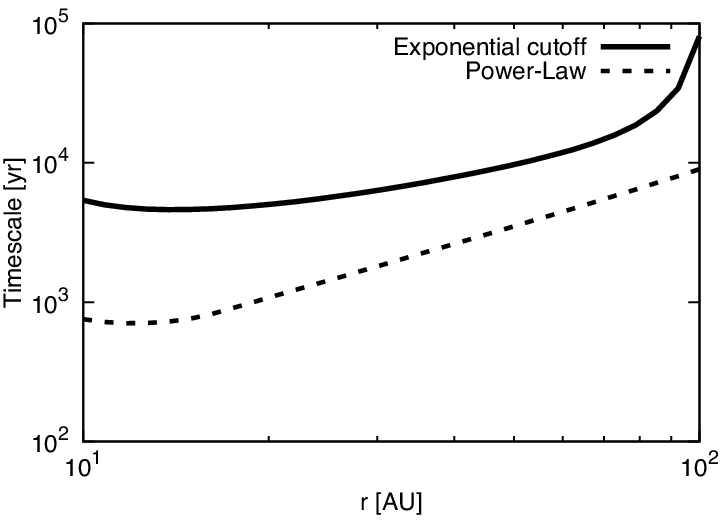}
 \caption{
The distribution of the most unstable wavelength (upper panel) and the growth timescale (lower panel) for exponentially cutoff disk model (solid line) and power-law disk model (dotted line) with $\alpha = 3\times 10^{-4} (r/100\,[\rm AU])^3$. The secular GI grows in the whole disk.
}
\label{fig:lambda_tgrow_alpha3}
\end{figure}

Fig. \ref{fig:lambda_tgrow_alpha2} shows the most unstable wavelength and growth timescale for exponentially cutoff and power-law disk model with $\alpha=\alpha_1 =3\times10^{-4}(r/100\,[\rm AU])^2$.
Fig. \ref{fig:lambda_tgrow_alpha2} shows that secular GI grows in the region $r\gtrsim 50$ AU, the growth timescale is a few $10^4$ yr and most unstable wavelength is about 10 AU for both models.
As a result, most of ring structures can be created by secular GI.
Moreover, in the case of exponentially cutoff disk model, secular GI grows even in the region around $\sim 40$ AU.
The most unstable wavelength is a few AU and smaller than the observed ring width.
We need detailed analysis of nonlinear growth of secular GI, but this result suggests that the rings whose width is a few AU are formed around $\sim 40$ AU and would be observed by future observations with higher spatial resolution.
Fig. \ref{fig:lambda_tgrow_alpha3} shows the radial distributions of the most unstable wavelength and the growth timescale with $\alpha =\alpha_2= 3\times 10^{-4}(r/100\,[\rm AU])^3$.
In this case, secular GI grows in the whole disk.
The growth timescale at the inner region ($r\lesssim 20$ AU) is about an order of magnitude smaller than that of the outer region ($r \sim 100$ AU).
Secular GI, hence, grows faster in the inner region than in the outer region.
The rapid growth of the secular GI in the inner region may cause the planet formation discussed in Section \ref{PlanetFormation}.

Although MRI turbulence in the midplane might be suppressed by electron heating, turbulent motion can be large in the upper region of the disk.
In this case, vertical structure of the disk may affect secular GI and changes the dispersion relation.
In this work, we use vertically integrated equations for linear stability analysis. 
To obtain more realistic dispersion relation we should take into account the vertical stratification of gas and dust components. 
In general it inevitably includes vertical shear of azimuthal velocity that causes Kelvin-Helmholtz instability (KHI). 
A detailed linear stability analysis of the KHI for two component system of gas and dust grains has been done by \cite{2006ApJ...641.1131M} that shows suppressed growth rates of KHI for larger dust grains. 
If the non-linear state of the KHI corresponds to a weak turbulence ($\alpha \lesssim 3\times 10^{-4} (r/100 {\rm [AU]})^3$), such effects of vertical stratification may not significantly affect the conclusion in this section. 
To clarify the actual outcome, however, we have to do three-dimensional non-linear simulations for such a multi-component system, which is beyond the scope of this paper. 

In this section, we discuss the growth of secular GI in the disk in which $\alpha$ changes with radius.
The region where secular GI is expected expands with decreasing $\alpha$.
When $\alpha$ is smaller than $\alpha_2$, the evolution of the disk does not change qualitatively.

\subsection{The Effect of Radial Drift of Dust}
\label{RadialDrift}
In this work, we adopt isothermal and uniform surface density as a background of the linear stability analysis.
As a result, the rotation velocity of gas and dust is the same in a steady state background.
In reality, however, the surface density and temperature are not uniform but depend on radius.
In general, the rotational velocity of the gas decreases because the gas is supported not only by the centrifugal force but also by the pressure gradient force against the gravity.
On the other hand, dust particles rotate with Keplerian velocity even in a nonuniform background.
As a consequence, the dust encounters the headwind, loses angular momentum, and drifts inward.
The drift timescale $t_{\rm dri} = r/v_{\rm dri}$ is $\sim 2\times 10^4$ yr for exponentially cutoff disk model and $\sim 10^4$ yr for power-law disk model.
These timescales are comparable to or smaller than the growth timescales of secular GI.
Therefore, the radial drift of dust will affect the growth of secular GI.
If the dust drift into the region where secular GI cannot grow before sufficient growth of the perturbations, it is difficult to explain the ring structure formation with secular GI.
If the disk extends larger than 100 AU, secular GI may grow with the radial drift in the outer region of the disk.
In this case, secular GI makes pressure bumps, and they prevent the fast radial drift and will support the further growth of secular GI.
The ring structures formed in the outer region ($ r \gtrsim 80 $ AU) move inward with the disk accretion.
If the rings are not destroyed in the secular GI stable region, ring structures can be formed by secular-GI even in the model that secular GI grows only in the outer region shown in Section \ref{comparison}.
This scenario naturally explains the observational result that the separations of all the rings appear to be comparable to the most unstable wavelength of secular GI around the disk radius of 100 AU.
This scenario may predict the existence of the extended outer disk with a much fainter ring-like structure, which can be tested by future ALMA observations.  
If the dust radial drift is strong enough, we should also consider the  effect of streaming instability \cite[SI,][]{2007ApJ...662..613Y,2005ApJ...620..459Y,2007ApJ...662..627J}.
Even if the non-linear growth of SI results in turbulence, the degree of turbulent mixing should be smaller than $\alpha = 3\times10^{-4}$,
because the value of $\alpha = 3\times10^{-4}$ is inferred from the observation of HL Tau.
Therefore, in the case of HL Tau, SI does not change our conclusion drastically.

Other mechanisms to create ring-like structures have been proposed after the observation of HL Tau.  
Actually some of them are not mutually exclusive but may possibly operate with secular GI. 
\cite{2016ApJ...821...82O} have proposed ring structure formation  by aggregate sintering.
Because sintering suppresses dust growth, the dust radius gets smaller in the sintering zone.
As a result, the radial drift velocity is small in the sintering zone and dust piles up there.
Obviously the dust condensation and piling up may positively affect the growth of secular GI. 
Dust condensation due to sintering can support the growth of secular GI.
Therefore, the ring structure of HL Tau may be formed by both secular GI and aggregate sintering.
\cite{2015IAUGA..2256118Z} have proposed the ring structure formation by dust growth around the condensation front.
They have not taken into account the radial drift of dust, but the dust condensation may occur in their scenario in the same manner as in \cite{2016ApJ...821...82O}.
Therefore, this also possibly supports the present scenario. 
The numerical simulation of secular GI that include these effects may enable the analysis on the interplay between these mechanisms, which will be our future work.

\subsection{Planet Formation by Secular GI}
\label{PlanetFormation}
\begin{figure}[t]
\includegraphics[width=8cm]{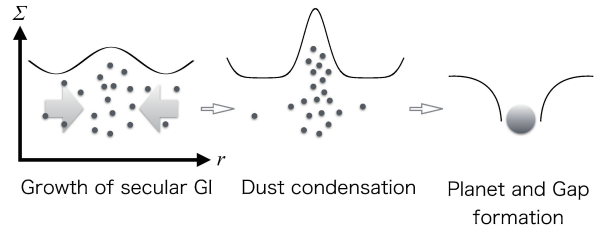}
\vspace{-2cm}
 \caption{Schematic picture of inner gap (D1 and D2 in \cite{2015ApJ...808L...3A}) formation by secular GI.
 Dust condensation occurs in the ring structures by secular GI.
 The dust grows in the ring and results in the formation of a planet. Finally, the planet opens the gap in the inner region.
}
\label{fig:Gapformation}
\end{figure}
Since secular GI concentrates the dust, planetesimals could be formed in rings.
Thus, we expect that planet formation might be promoted as a result of secular GI.
The growth timescales of secular GI in the outer regions ($\sim 100$AU) are comparable to the estimated age of the central star and the disk.
 Therefore natural interpretation is that the outer region is in the early growth phase of secular GI. 
In contrast to the outer region, however, the inner region may have gone to the non-linear stages of secular GI because the growth timescales of secular GI in the inner regions are smaller than those in the outer regions.
Therefore, we speculate that the non-linear growth of secular GI may have resulted in the formation of planets that make gaps observed at $r\sim 10$ AU and $\sim 30$ AU (Fig. \ref{fig:Gapformation}).
The amounts of missing dust in the inner two gaps are estimated by \cite{2016ApJ...816...25P}:  $\sim 7.2M_\oplus$ for the inner-most gap (D1 in \cite{2015ApJ...808L...3A}) and $\sim 21.6 M_\oplus$ for the next gap (D2).
We can simply expect the missing dust mass in the gap corresponds to the mass of heavy elements in the hypothetical planet in the gap. 
Since we adopt the dust-to-gas mass ratio $\epsilon =0.02$ for our disk model that causes the secular GI, we should expect that the actual total mass of the planet in the gap can be about 50 times larger than the heavy element mass of the planet, if all the gaseous components of the disk in the gap accrete onto the planet. 
This argument means that we set the expected ranges of masses of the planets in the inner two gaps are $7.2 M_\oplus \-- 1.1 M_{\rm J}$ and $21.6M_\oplus \-- 3.4 M_{\rm J}$, for D1 and D2, respectively.

When dust grains grow sufficiently in a high density region, the opacity of the ring decreases and the dust thermal radiation may remain weak in the ring. 
Even after the formation of planetesimals or protoplanets, the ring may remain as an invisible planetesimal belt. 
In this scenario, we expect to have a chance to observe the outcome of planet formation only after the mass of a planet increases sufficiently. 

   On the other hand, various authors estimated the masses of the hypothetical planets in the gaps by assuming the structure in the disk is created by hypothetical planets in the gap. 
For example, \cite{2015ApJ...806L..15K} provided an approximate relation of the gap depth/width and the planet mass. 
According to their relation we can obtain the hypothetical planet mass in the gap: $\sim 0.6M_{\rm J}$ for D1 and $\sim 1M_{\rm J}$ for D2. 
This may mean that gaseous components in the disk do not perfectly accrete onto the planet in the gap. 
Despite the simplification in our argument, the-order-of-magnitude agreement between the maximum planet mass expected from the formation mechanism and the mass estimated from the resultant gap structure is reasonable and may motivate further investigation in this line of research. 

One of the differences between the present paper and other papers on HL Tau is in the connection of planet formation mechanism to the ring structure in the disk. 
Papers on gap opening dynamics simply assumed the existence of planets and did not discussed their origins \cite[e.g.,][]{2015ApJ...806L..15K,2015MNRAS.453L..73D,2016ApJ...818...76J}.
In contrast, other papers on the mechanism to create multiple rings without embedded planets did not directly link the ring formation to mechanisms of planet formation \cite[e.g.,][]{2015IAUGA..2256118Z,2016ApJ...821...82O} partly because the ring formation in the disk is regarded as a transient phenomenon \cite[]{2016ApJ...821...82O}. 
To theoretically verify the discussion in this section, we need to investigate the nonlinear growth of secular GI in the realistic situation taking into account the dust growth and radial drift. 

\section{SUMMARY}
In this work, we perform linear stability analyses of secular GI adopting the physical values obtained from observation of HL Tau as background state.
We calculated the radial distributions of the most unstable wavelength and the growth timescale of secular GI and verify that secular GI can form the ring structures observed in HL Tau.
We use two background states: exponentially cutoff disk model and power-law disk model.
Both models give similar results.
We showed that secular GI grows in the outer region $r\gtrsim 80$ AU when we adopt the dust surface density, temperature, and a turbulent viscosity coefficient obtained from the observation.
Around a radius of 100AU, the most unstable wavelength is about 10 AU, and growth timescale is about $10^5$ yr and $10^4$ yr for exponentially cutoff and power-law disk model respectively.
The wavelength is consistent with the observed separations of the rings.
When we adopt a turbulent viscosity coefficient $\alpha =3\times 10^{-4}(r/100\, [\rm AU])^2$, secular GI grows in the region $r \gtrsim 50 $ AU. 
Thus, secular GI forms all rings whose width is about 10 AU.
Moreover, in the case that we adopt a turbulent viscosity coefficient $\alpha =3\times 10^{-4}(r/100\, [\rm AU])^3$, secular GI grows in the whole disk. 
Since the growth timescale in the inner region $r\lesssim 30$ AU is about an order of magnitude smaller than that of in the outer region $r\sim 100$ AU, secular GI can rapidly grow in the inner region.
If the rings formed in the outer region move inward with the disk accretion, secular GI can naturally form the equi-separation rings if the separations remain the same in the accretion process. 

Since secular GI concentrate the dust, it supports the dust growth.
Because the growth timescale of secular GI in the inner region may be smaller than the estimated age of HL Tau,
plant could be formed in rings in the inner region.
In this case, resulting objects are expected to create the gaps at $r\sim 10$ AU and $\sim $ 30 AU.
If the scenario described in this article is indeed realized in HL Tau, it means that we are witnessing both the early stage of planet formation and the stage after the planet formation: early growth of the most unstable mode of secular GI and the gap structure carved by resulting planets. 
Therefore, further observation of HL Tau would give more detailed information for various stages of planet formation.
\label{conclusion}

SI thanks Neal Turner, Andrew Youdin, Anders Johansen, and Wladimir Lyra
for visiting him and discussing on various issues for protoplanetary disks.
The authors also thank Hiroshi Kobayashi, Takayuki Muto, 
Misato Fukagawa, Munetake Momose, Kazuyuki Omukai, and Shigeo Kimura.
This work was partly supported by a Grant-in-Aid for JSPS Fellows and JSPS KAKENHI Grant Numbers 23244027 and 23103005.

\bibliographystyle{apj}

\end{document}